\documentclass[twocolumn,prl,showpacs,superscriptaddress]{revtex4}%
\usepackage{amssymb}
\usepackage{amsmath}
\usepackage{amsfonts}
\usepackage{graphicx}%
\begin{document}
\title{All-Versus-Nothing Violation of Local Realism for Two Entangled Photons}
\author{Zeng-Bing Chen}
\affiliation{Department of Modern Physics, University of Science and Technology of China,
Hefei, Anhui 230027, China}
\author{Jian-Wei Pan}
\affiliation{Institut f\"{u}r Experimentalphysik, Universit\"{a}t Wien, Boltzmanngasse 5,
1090 Wien, Austria}
\author{Yong-De Zhang}
\affiliation{Department of Modern Physics, University of Science and Technology of China,
Hefei, Anhui 230027, China}
\author{\v{C}aslav Brukner}
\affiliation{Institut f\"{u}r Experimentalphysik, Universit\"{a}t Wien, Boltzmanngasse 5,
1090 Wien, Austria}
\author{Anton Zeilinger}
\affiliation{Institut f\"{u}r Experimentalphysik, Universit\"{a}t Wien, Boltzmanngasse 5,
1090 Wien, Austria}

\begin{abstract}
It is shown that the Greenberger-Horne-Zeilinger theorem can be generalized to
the case with only two entangled particles. The reasoning makes use of two
photons which are maximally entangled both in polarization and in spatial
degrees of freedom. In contrast to Cabello's argument of \textquotedblleft all
versus nothing\textquotedblright\ nonlocality with four photons [Phys. Rev.
Lett. \textbf{87}, 010403 (2001)], our proposal to test the theorem can be
implemented with linear optics and thus is well within the reach of current
experimental technology.

\end{abstract}
\maketitle

Bell's theorem \cite{Bell}, which is derived from Einstein, Podolsky, and
Rosen's (EPR's) notion of local realism \cite{EPR}, represents the most
radical departure of quantum mechanics (QM) from one's classical intuitions.
On the one hand, Bell's inequalities (BI) state that certain statistical
correlations predicted by QM for measurements on two-particle
\textit{ensembles} cannot be understood within a realistic picture based on
local properties of each individual particle. On the other hand, an
unstatisfactory feature in the derivation of BI is that such a local realistic
and thus classical picture can explain perfect correlations and is only in
conflict with statistical prediction of the theory.

Strikingly, \textquotedblleft Bell's theorem without
inequalities\textquotedblright\ has been demonstrated for multiparticle
Greenberger-Horne-Zeilinger (GHZ) states \cite{GHZ-89,GHZ-90,Pan-GHZ}, where
the contradiction between QM and local realistic theories arises even for
\textit{definite predictions}. The quantum nonlocality can thus, in principle,
be manifest in a single run of a certain measurement. This is known as the
\textquotedblleft all versus nothing\textquotedblright\ proof of Bell's
theorem. In addition, the GHZ contradiction applies for all ($100\%$)
multiparticle systems that are in the same GHZ state. In the sense that it is
for definite predictions and for all systems the GHZ theorem represents the
strongest conflict between QM and local realism. However, the GHZ reasoning
requires at least three particles and, consequently, three space-like
separated regions (observers). This can be seen as a sort of three-particle
quantum nonlocality, which differs from the two-particle quantum nonlocality
as implied in usual BI.

Then Hardy's argument of ``quantum nonlocality without inequalities''\ for
nonmaximally entangled biparticle states \cite{Hardy} came as a surprise. Now
it is known as ``the best version of Bell's theorem''\ \cite{Mermin} for
two-dimensional two-particle systems. However, compared to the GHZ case, in
Hardy's proof only a fraction ($\lesssim9\%$) of the photon pairs shows a
contradiction with local realism. Most recently, another way to reveal sharper
violations of local realism for two-particle entangled states in
higher-dimensional Hilbert spaces was found \cite{Zeilinger-n,n-dim}. For the
two-particle entangled states of high-dimensionality, the violation of local
realism has more resistance to noise, but is still statistical. Motivated by
Hardy and the high-dimensional versions of Bell's theorem, one may ask: Can
the conflict between QM and local realism arise even for the definite
predictions and for all $(100\%)$ of the photon pairs in the same entangled state?

In this Letter we answer the question affirmatively by demonstrating
\textit{an all-versus-nothing nonlocality for two photons which are maximally
entangled both in polarization and in spatial (path) degrees of freedom}. Such
a \textquotedblleft double entanglement\textquotedblright\ plays a crucial
role in our demonstration. From a formal aspect, our demonstration is a
further development of Cabello's \cite{Cabello-a,Cabello-b} elegant proof of
quantum nonlocality without inequalities for two observers who possess two
pairs of maximally entangled qubits, i.e., four two-level particles. Lvovsky
demonstrated that the nonlinear optics at a single-photon level is required
for a demonstration of Cabello's quantum nonlocality without implicit
assumption of noncontextuality \cite{Lvovsky}. Unfortunately, such an
experimental test of Cabello's quantum nonlocality is beyond the present level
of quantum optical technology. By contrast, the experiment proposed here has
two advantanges over Cabello's proposal. First, the observers need to possess
only one pair of entangled photons at a time and, second, it can readily be
done as it needs only linear optics elements.

Currently, the most widely used reliable source of polarization-entangled
photons is parametric down-conversion in a nonlinear optical crystal
\cite{Kwiat}. Here we need two-photon states that are maximally entangled both
in polarization and in path degrees of freedom. Figure $1$ shows the setup
\cite{Herzog,Simon-Pan} for generating pairs of polarization and path
entangled photons. A pump pulse passing through the crystal can create, with a
small probability, entangled pairs of photons in the spatial (path) modes
$d_{1}$ and $u_{2}$. For definiteness, we assume that the entangled photon
pairs are in the polarization-entangled state $\left\vert \Psi^{-}%
\right\rangle _{12}=\frac{1}{\sqrt{2}}(\left\vert H\right\rangle
_{1}\left\vert V\right\rangle _{2}-\left\vert V\right\rangle _{1}\left\vert
H\right\rangle _{2})$, where $\left\vert H\right\rangle $ ($\left\vert
V\right\rangle $) stands for photons with horizontal (vertical) polarization.
Now if the pump is reflected through the crystal a second time, then there is
another possibility for producing entangled pairs of photons into the path
modes $u_{1}$ and $d_{2}$ that are opposite to the first modes $d_{1}$ and
$u_{2}$. The two possible ways of producing the entangled photon pairs may
interfere \cite{Herzog}. By properly adjusting the distance between the mirror
and the crystal, the setup in Fig. $1$ generates the doubly entangled
two-photon state \cite{Simon-Pan}
\begin{equation}
\left\vert \Psi\right\rangle _{12}=\frac{1}{2}(\left\vert H\right\rangle
_{1}\left\vert V\right\rangle _{2}-\left\vert V\right\rangle _{1}\left\vert
H\right\rangle _{2})(\left\vert u\right\rangle _{1}\left\vert d\right\rangle
_{2}-\left\vert d\right\rangle _{1}\left\vert u\right\rangle _{2}), \label{pp}%
\end{equation}
which is just the desired state entangled maximally both in polarization and
in path. Here photon-$1$ and photon-$2$ are, respectively, possessed by two
observers, Alice and Bob, who are space-like separated; $\left\vert
u\right\rangle $ and $\left\vert d\right\rangle $ denote two orthonormal path
states of photons. With emphasis, we note that the state (\ref{pp}) indeed
corresponds to the case where there is one and only one pair production after
the pump pulse passes through the BBO crystal twice.%
\begin{figure}
[ptb]
\begin{center}
\includegraphics[
height=0.9406in,
width=2.2183in
]%
{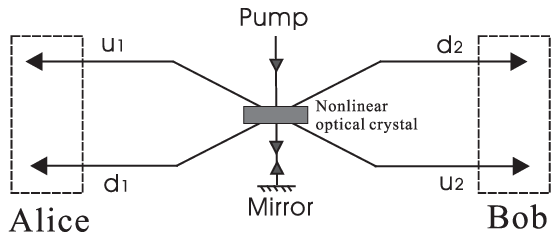}%
\caption{Setup for generating pairs of photons entangled both in polarization
and path.}%
\label{fig1}%
\end{center}
\end{figure}

One can define the following Pauli-type operators for both the polarization
and the path degrees of freedom:
\begin{align}
\sigma_{x}  &  =\left\vert H\right\rangle \left\langle V\right\vert
+\left\vert V\right\rangle \left\langle H\right\vert ,\;\;\sigma
_{z}=\left\vert H\right\rangle \left\langle H\right\vert -\left\vert
V\right\rangle \left\langle V\right\vert ;\nonumber\\
\sigma_{x}^{\prime}  &  =\left\vert u\right\rangle \left\langle d\right\vert
+\left\vert d\right\rangle \left\langle u\right\vert ,\;\;\sigma_{z}^{\prime
}=\left\vert u\right\rangle \left\langle u\right\vert -\left\vert
d\right\rangle \left\langle d\right\vert . \label{pauli}%
\end{align}
For convenience and clarity, in the following we denote $z_{i}=\sigma_{zi}$,
$x_{i}=\sigma_{xi}$, $z_{i}^{\prime}=\sigma_{zi}^{\prime}$, ${x}_{i}^{\prime
}=\sigma_{xi}^{\prime}$ ($i=1,2$) and, use $(\cdot)$ to separate operators or
operator products that can be identified as EPR's local \textquotedblleft
elements of reality\textquotedblright. Then one can easily check the following
eigenequations
\begin{align}
&  \ \left.  z_{1}\cdot z_{2}\left\vert \Psi\right\rangle _{12}=-\left\vert
\Psi\right\rangle _{12},\;z_{1}^{\prime}\cdot z_{2}^{\prime}\left\vert
\Psi\right\rangle _{12}=-\left\vert \Psi\right\rangle _{12},\right.
\label{e1}\\
&  \ \left.  x_{1}\cdot x_{2}\left\vert \Psi\right\rangle _{12}=-\left\vert
\Psi\right\rangle _{12},\;x_{1}^{\prime}\cdot x_{2}^{\prime}\left\vert
\Psi\right\rangle _{12}=-\left\vert \Psi\right\rangle _{12},\right.
\label{e3}\\
&  \ \left.  z_{1}z_{1}^{\prime}\cdot z_{2}\cdot z_{2}^{\prime}\left\vert
\Psi\right\rangle _{12}=\left\vert \Psi\right\rangle _{12},\right.
\label{e5}\\
&  \ \left.  x_{1}x_{1}^{\prime}\cdot x_{2}\cdot x_{2}^{\prime}\left\vert
\Psi\right\rangle _{12}=\left\vert \Psi\right\rangle _{12},\right.
\label{e6}\\
&  \ \left.  z_{1}\cdot x_{1}^{\prime}\cdot z_{2}x_{2}^{\prime}\left\vert
\Psi\right\rangle _{12}=\left\vert \Psi\right\rangle _{12},\right.
\label{e7}\\
&  \ \left.  x_{1}\cdot z_{1}^{\prime}\cdot x_{2}z_{2}^{\prime}\left\vert
\Psi\right\rangle _{12}=\left\vert \Psi\right\rangle _{12},\right.
\label{e8}\\
&  \ \left.  z_{1}z_{1}^{\prime}\cdot x_{1}x_{1}^{\prime}\cdot z_{2}%
x_{2}^{\prime}\cdot x_{2}z_{2}^{\prime}\left\vert \Psi\right\rangle
_{12}=-\left\vert \Psi\right\rangle _{12}.\right.  \label{e9}%
\end{align}
Equations (\ref{e1})-(\ref{e9}) contain only local operators, i.e., ($z_{1}$,
$z_{1}^{\prime}$, $x_{1}$, $x_{1}^{\prime}$, $z_{1}z_{1}^{\prime}$,
$x_{1}x_{1}^{\prime}$, $z_{1}\cdot x_{1}^{\prime}$, $x_{1}\cdot z_{1}^{\prime
}$, and $z_{1}z_{1}^{\prime}\cdot x_{1}x_{1}^{\prime}$) for Alice and ($z_{2}%
$, $z_{2}^{\prime}$, $x_{2}$, $x_{2}^{\prime}$, $z_{2}\cdot z_{2}^{\prime}$,
$x_{2}\cdot x_{2}^{\prime}$, $z_{2}x_{2}^{\prime}$, $x_{2}z_{2}^{\prime}$, and
$z_{2}x_{2}^{\prime}\cdot x_{2}z_{2}^{\prime}$) for Bob. In particular, Eqs.
(\ref{e1})-(\ref{e9}) allow Alice (Bob) to assign values with certainty to
Bob's local operators $z_{2}$, $z_{2}^{\prime}$, $x_{2}$, $x_{2}^{\prime}$,
$z_{2}x_{2}^{\prime}$, and $x_{2}z_{2}^{\prime}$ (Alice's local operators
$z_{1}$, $z_{1}^{\prime}$, $x_{1}$, $x_{1}^{\prime}$, $z_{1}z_{1}^{\prime}$,
and $x_{1}x_{1}^{\prime}$) by measuring her local observables (his local
observables) without in any way disturbing Bob's (Alice's) photon. It is the
idea of EPR's criterion of elements of reality to establish a local realistic
interpretation of the quantum-mechanical results (\ref{e1})-(\ref{e9}) by
assuming that the individual value of any operator ($z_{1}$, $z_{1}^{\prime}$,
$x_{1}$, $x_{1}^{\prime}$, $z_{1}z_{1}^{\prime}$, and $x_{1}x_{1}^{\prime}$)
at Alice's side and ($z_{2}$, $z_{2}^{\prime}$, $x_{2}$, $x_{2}^{\prime}$,
$z_{2}x_{2}^{\prime}$, and $x_{2}z_{2}^{\prime}$) at Bob's side is
predetermined. These predetermined values are denoted by $v\left(
z_{i}\right)  $, $v\left(  z_{i}^{\prime}\right)  $, $v\left(  x_{i}\right)
$, $v\left(  x_{i}^{\prime}\right)  $, $v(z_{1}z_{1}^{\prime})$, $v(x_{1}%
x_{1}^{\prime})$, $v\left(  z_{2}x_{2}^{\prime}\right)  $, and $v\left(
x_{2}z_{2}^{\prime}\right)  $ with $v=\pm1$. To be consistent with Eqs.
(\ref{e1})-(\ref{e9}), local realistic theories thus predict
\begin{align}
&  \ \left.  v(z_{1})v(z_{2})=-1,\;\;\;\;v(z_{1}^{\prime})v(z_{2}^{\prime
})=-1,\right. \label{le2}\\
&  \ \left.  v\left(  x_{1}\right)  v\left(  x_{2}\right)
=-1,\;\;\;\;v\left(  x_{1}^{\prime}\right)  v\left(  x_{2}^{\prime}\right)
=-1,\right. \label{le4}\\
&  \ \left.  v\left(  z_{1}z_{1}^{\prime}\right)  v\left(  z_{2}\right)
v\left(  z_{2}^{\prime}\right)  =1,\;v\left(  x_{1}x_{1}^{\prime}\right)
v\left(  x_{2}\right)  v\left(  x_{2}^{\prime}\right)  =1,\right.
\label{le6}\\
&  \ \left.  v\left(  z_{1}\right)  v\left(  x_{1}^{\prime}\right)  v\left(
z_{2}x_{2}^{\prime}\right)  =1,\;v\left(  x_{1}\right)  v\left(  z_{1}%
^{\prime}\right)  v\left(  x_{2}z_{2}^{\prime}\right)  =1,\right.
\label{le8}\\
&  \ \left.  v\left(  z_{1}z_{1}^{\prime}\right)  v\left(  x_{1}x_{1}^{\prime
}\right)  v\left(  z_{2}x_{2}^{\prime}\right)  v\left(  x_{2}z_{2}^{\prime
}\right)  =-1.\right.  \label{le9}%
\end{align}
But in fact, Eqs. (\ref{le2})-(\ref{le9}) are mutually inconsistent:
Multiplying Eqs. (\ref{le2})-(\ref{le8}), one gets $v\left(  z_{1}%
z_{1}^{\prime}\right)  v\left(  x_{1}x_{1}^{\prime}\right)  v\left(
z_{2}x_{2}^{\prime}\right)  v\left(  x_{2}z_{2}^{\prime}\right)  =1$ due to
the fact that $v^{2}\left(  z_{i}\right)  =v^{2}\left(  z_{i}^{\prime}\right)
=v^{2}\left(  x_{i}\right)  =v^{2}\left(  x_{i}^{\prime}\right)  =1$, and this
is then in conflict with Eq. (\ref{le9}). Thus, the quantum-mechanical
predictions (\ref{e1})-(\ref{e9}) are incompatible with the ones imposed by
local realistic theories. The contradiction between QM and local realism
occurs for definite predictions and for all ($100\%$) of the photon pairs.
This completes the demonstration of an all-versus-nothing nonlocality for our
two-photon case.

An important point deserves further comment. It is well known that the
original GHZ argument needs at least three spatially separated particles in
order to establish the properties used in the argument as EPR's elements of
reality. Therefore the question arises whether it is also possible to achieve
the same in a two-particle situation as suggested in this paper. We are able
to achieve this goal for two reasons: First, the number of variables used in
the argument is enlarged compared to the original GHZ argument. Second, and
most importantly, the nine variables can be arranged in three groups of three
each, where the three variables of each group are measured by one and the same
apparatus when establishing them as ERP's elements of reality, as we will show
below. This eliminates the necessity of an argument based on counterfactuality
as it is not necessary to assume any of these variables to be independent of
experimental context.

Actually, the above argument can be understood from another perspective. By
defining $\left\vert H\right\rangle \left\vert u\right\rangle \equiv\left\vert
0\right\rangle $, $\left\vert H\right\rangle \left\vert d\right\rangle
\equiv\left\vert 1\right\rangle $, $\left\vert V\right\rangle \left\vert
u\right\rangle \equiv\left\vert 2\right\rangle $ and $\left\vert
V\right\rangle \left\vert d\right\rangle \equiv\left\vert 3\right\rangle $,
$\left\vert \Psi\right\rangle _{12}$ can be rewritten as $\left\vert
\Psi\right\rangle _{12}=\frac{1}{2}(\left\vert 0\right\rangle _{1}\left\vert
3\right\rangle _{2}-\left\vert 1\right\rangle _{1}\left\vert 2\right\rangle
_{2}-\left\vert 2\right\rangle _{1}\left\vert 1\right\rangle _{2}+\left\vert
3\right\rangle _{1}\left\vert 0\right\rangle _{2})$, which is, in fact, a
two-particle maximally-entangled state in a four-dimensional Hilbert space.
This then implies that the\ GHZ-type argument has been indeed generalized to
the case with only two entangled four-level particles. In contrast to the
original GHZ proposal, our scheme requires only two space-like separated regions.

In a real experiment, the perfect correlations and ideal measurement devices
are practically impossible. To face this difficulty, a Bell-Mermin inequality
for $\left\vert \Psi\right\rangle _{12}$ is desirable. Similarly to Ref.
\cite{Cabello-b}, one can introduce the operator $\mathcal{O}=-z_{1}\cdot
z_{2}-z_{1}^{\prime}\cdot z_{2}^{\prime}-x_{1}\cdot x_{2}-x_{1}^{\prime}\cdot
x_{2}^{\prime}+z_{1}z_{1}^{\prime}\cdot z_{2}\cdot z_{2}^{\prime}+x_{1}%
x_{1}^{\prime}\cdot x_{2}\cdot x_{2}^{\prime}+z_{1}\cdot x_{1}^{\prime}\cdot
z_{2}x_{2}^{\prime}+x_{1}\cdot z_{1}^{\prime}\cdot x_{2}z_{2}^{\prime}%
-z_{1}z_{1}^{\prime}\cdot x_{1}x_{1}^{\prime}\cdot z_{2}x_{2}^{\prime}\cdot
x_{2}z_{2}^{\prime}$. It can be directly seen from Eqs. (\ref{e1})-(\ref{e9})
that $\mathcal{O}$ satisfies
\begin{equation}
\mathcal{O}\left\vert \Psi\right\rangle _{12}=9\left\vert \Psi\right\rangle
_{12}. \label{oe}%
\end{equation}
However, following Ref. \cite{Cabello-b} local realistic theories predict the
observed values of $\mathcal{O}$
\begin{equation}
\left\langle \mathcal{O}\right\rangle _{LRT}\leq7, \label{o7}%
\end{equation}
which is in contradiction with the quantum mechanical prediction (\ref{oe}).
For observing the violation of the inequality (\ref{o7}), one needs the
doubly-entangled state with a visibility better than $7/9\approx77.8\%$. Here,
we would like to mention that BI for \textquotedblleft
qudits\textquotedblright\ have more resistance to noise and the required
visibility can be reduced to about $69\%$ for four-dimensional systems
\cite{Zeilinger-n,n-dim}.

Though the above argument is formally similar to the reasoning of Cabello's
theorem \cite{Cabello-a,Cabello-b}, at this stage the advantages of our scheme
are already manifest. Our argument works for \textit{two} entangled photons,
whose path and polarization degrees of freedom are used. Experimentally,
manipulating a single pair of entangled photons is much easier than
manipulating two pairs. These features are essential for an experimental test
of the GHZ-type theorem proposed here.

We now further discuss the noncontextuality issue to validate our
all-versus-nothing quantum nonlocality argument. In the argument the same
operators may appear in different equations (\ref{e1})-(\ref{e9}). For
example, $z_{1}z_{1}^{\prime}$ and $x_{1}x_{1}^{\prime}$ not only appear
separately in Eqs. (\ref{e5}) and (\ref{e6}), but also appear jointly in Eq.
(\ref{e9}). In order for the argument to hold, it is, however, necessary to
assign always a single value to the same operator, though it can appear in
different equations. Therefore, one either has to assume noncontexuality
(e.g., measurement of $z_{1}z_{1}^{\prime}$ does not disturb the value of
$x_{1}x_{1}^{\prime}$ and vice versa) or one has to be able to measure
$z_{1}z_{1}^{\prime}$, $x_{1}x_{1}^{\prime}$ and $z_{1}z_{1}^{\prime}\cdot
x_{1}x_{1}^{\prime}$ with the same apparatuses. Lvovsky noticed that this
would then require quantum controlled-NOT (CNOT) operation to apply on all
($100\%$) photon pairs to demonstrate Cabello's quantum nonlocality in the
original proposal with two photon pairs. Unfortunately, this requires
nonlinear optics.

This CNOT operation is equivalent to making a complete Bell-state
discrimination (see Refs. \cite{Cabello-a,Cabello-b,Lvovsky}), in which the
Bell states are $\left\vert \Psi^{\pm}\right\rangle =\frac{1}{\sqrt{2}%
}(\left\vert H\right\rangle \left\vert V\right\rangle \pm\left\vert
V\right\rangle \left\vert H\right\rangle )$ and $\left\vert \Phi^{\pm
}\right\rangle =\frac{1}{\sqrt{2}}(\left\vert H\right\rangle \left\vert
H\right\rangle \pm\left\vert V\right\rangle \left\vert V\right\rangle )$. It
has been well known initially in the context of quantum teleportation
\cite{Bennett93,Dik-Pan,teleport-PRL} that such a full Bell-state measurement
is impossible with only linear optics and necessitates nonlinear optical
interactions at a single-photon level \cite{impossible}, which is very
challenging experimentally. Thus, an experimental test of Cabello's
nonlocality cannot be achieved by existing technology \cite{Lvovsky}.

However, within our two-photon proposal, the above problem does not exist
since quantum CNOT operations can be easily implemented between two different
degrees of freedom of single photons. Actually, measuring $z_{1}z_{1}^{\prime
}\cdot x_{1}x_{1}^{\prime}$ in the present scheme is equivalent to performing
a complete Bell-state measurement, with the four Bell states
\begin{align}
\left\vert \psi^{\pm}\right\rangle  &  =\frac{1}{\sqrt{2}}(\left\vert
H\right\rangle \left\vert d\right\rangle \pm\left\vert V\right\rangle
\left\vert u\right\rangle ),\nonumber\\
\left\vert \phi^{\pm}\right\rangle  &  =\frac{1}{\sqrt{2}}(\left\vert
H\right\rangle \left\vert u\right\rangle \pm\left\vert V\right\rangle
\left\vert d\right\rangle ), \label{bspp}%
\end{align}
instead of $\left\vert \Psi^{\pm}\right\rangle $ and $\left\vert \Phi^{\pm
}\right\rangle $. The complete discrimination of the four Bell states in Eq.
(\ref{bspp}) has been realized in the \textquotedblleft two-particle
analog\textquotedblright\ of the quantum teleportation experiment performed in
Rome \cite{teleport-PRL}. Such a complete Bell-state discrimination can be
accomplished with linear optics and almost $100\%$ efficiency.

Thus the difficulty of measuring simultaneously $z_{1}z_{1}^{\prime}$,
$x_{1}x_{1}^{\prime}$ and $z_{1}z_{1}^{\prime}\cdot x_{1}x_{1}^{\prime}$ has
been eliminated by our two-photon proposal. We now consider the question of
how to measure the quantities such as $z_{1}$, $x_{1}^{\prime}$ and
$z_{1}\cdot x_{1}^{\prime}$ in our two-photon scheme. It is obvious that the
measurements of $z_{1}$, $x_{1}^{\prime}$ and $z_{1}\cdot x_{1}^{\prime}$ have
to be performed on the single photon possessed by Alice. Therefore, in order
to avoid the noncontexuality assumption, one must design an apparatus such
that it can give the measurement results of $z_{1}$, $x_{1}^{\prime}$ and
$z_{1}\cdot x_{1}^{\prime}$ simultaneously.

Similar considerations would thus lead to the following six apparatuses
(similar apparatuses have been proposed by Simon \textit{et al}. \cite{Simon}
in a different context), which are sufficient to solve the problem just
mentioned. Apparatus-$1$ measures $z_{1}$, $x_{1}^{\prime}$ and $z_{1}\cdot
x_{1}^{\prime}$; apparatus-$2$ measures $x_{2}$, $x_{2}^{\prime}$ and
$x_{2}\cdot x_{2}^{\prime}$; apparatus-$3$ measures $z_{1}^{\prime}$, $x_{1}$
and $x_{1}\cdot z_{1}^{\prime}$; apparatus-$4$ measures $z_{2}$,
$z_{2}^{\prime}$ and $z_{2}\cdot z_{2}^{\prime}$; apparatus-$5$ measures
$z_{1}z_{1}^{\prime}$, $x_{1}x_{1}^{\prime}$ and $z_{1}z_{1}^{\prime}\cdot
x_{1}x_{1}^{\prime}$; apparatus-$6$ measures $z_{2}x_{2}^{\prime}$,
$x_{2}z_{2}^{\prime}$ and $z_{2}x_{2}^{\prime}\cdot x_{2}z_{2}^{\prime}$.
Fortunately enough, each of these apparatuses measures different local
observables and, more importantly, the six apparatuses can be realized without
any mutual conflict.%
\begin{figure}
[ptb]
\begin{center}
\includegraphics[
height=3.4579in,
width=2.6667in
]%
{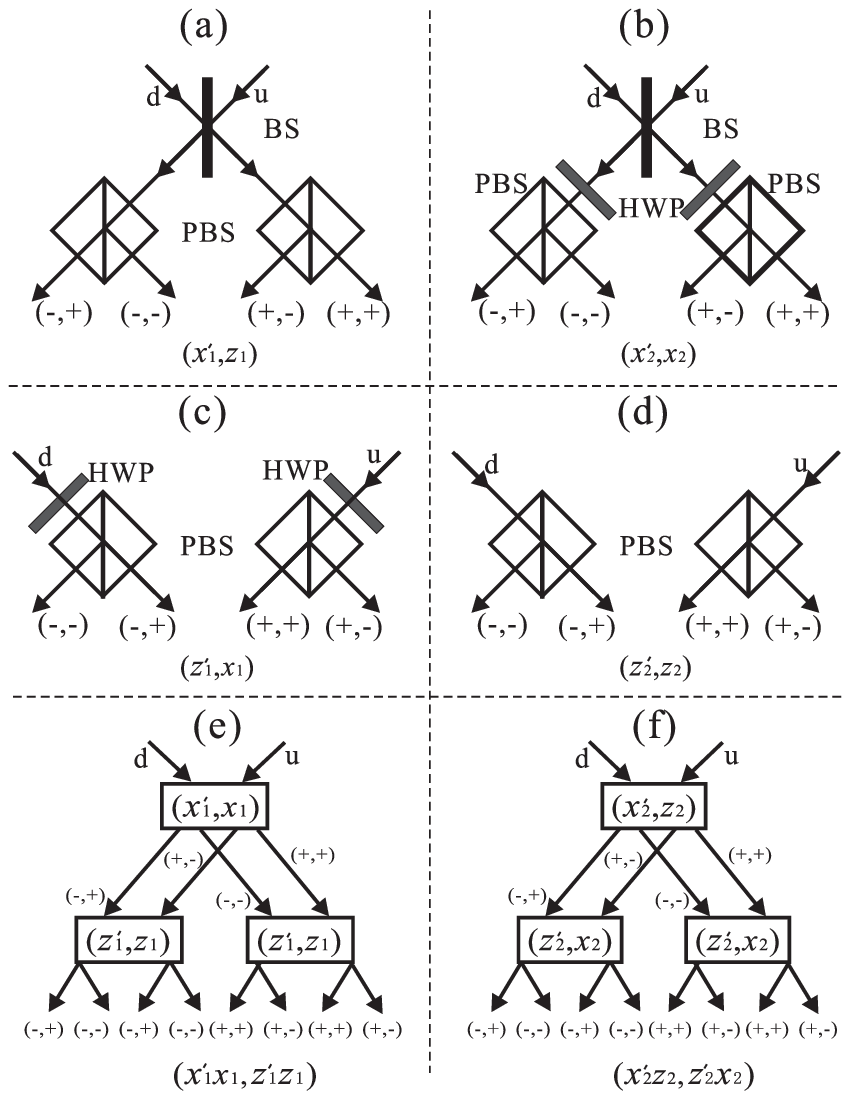}%
\caption{Six apparatuses for measuring $z_{1}$, $x_{1}^{\prime}$ and
$z_{1}\cdot x_{1}^{\prime}$ (a); $x_{2}$, $x_{2}^{\prime}$ and $x_{2}\cdot
x_{2}^{\prime}$ (b); $z_{1}^{\prime}$, $x_{1}$ and $x_{1}\cdot z_{1}^{\prime}$
(c); $z_{2}$, $z_{2}^{\prime}$ and $z_{2}\cdot z_{2}^{\prime}$ (d);
$z_{1}z_{1}^{\prime}$, $x_{1}x_{1}^{\prime}$ and $z_{1}z_{1}^{\prime}\cdot
x_{1}x_{1}^{\prime}$ (e); $z_{2}x_{2}^{\prime}$, $x_{2}z_{2}^{\prime}$ and
$z_{2}x_{2}^{\prime}\cdot x_{2}z_{2}^{\prime}$ (f). By $\pm$, we mean $\pm1$.}%
\label{fig2}%
\end{center}
\end{figure}

Figures $2$ (a-d) show the first four apparatuses, which require only simple
linear optical elements [e.g., the beam splitters (BS), polarizing beam
splitters (PBS) and half wave plates (HWP) rotated at $45^{\circ}$] and
single-photon detectors. Note that a PBS reflects $V$ photons and transmits
$H$ photons, and a BS (HWP) affects the following transformations: $\left\vert
u\right\rangle \rightarrow(\left\vert u\right\rangle +\left\vert
d\right\rangle )/\sqrt{2}$ and $\left\vert d\right\rangle \rightarrow
(\left\vert u\right\rangle -\left\vert d\right\rangle )/\sqrt{2}$ ($\left\vert
H\right\rangle \rightarrow(\left\vert H\right\rangle +\left\vert
V\right\rangle )/\sqrt{2}$ and $\left\vert V\right\rangle \rightarrow
(\left\vert H\right\rangle -\left\vert V\right\rangle )/\sqrt{2}$). Let us
consider, e.g., apparatus $1$. Since apparatus $1$ measures $z_{1}$ and
$x_{1}^{\prime}$ \textit{simultaneously}, it actually also gives the
measurement result of $z_{1}\cdot x_{1}^{\prime}$. Figures $2$(e) and $2$(f)
show the last two apparatuses, each of which is made up of three of the first
four apparatuses. Figure $2$(e) [$2$(f)] also shows that the information
obtained in the first stage of the measurement about the values of $x_{1}$ and
$x_{1}^{\prime}$ ($z_{2}$ and $x_{2}^{\prime}$) must be partially erased in
such a way that only information about the product $x_{1}x_{1}^{\prime}$
($z_{2}x_{2}^{\prime}$) is retained, to enable the measurement of $z_{1}%
z_{1}^{\prime}$ ($x_{2}z_{2}^{\prime}$) at the last stage \cite{Simon}.
Apparatus $5$ (apparatus $6$) measures $z_{1}z_{1}^{\prime}$ and $x_{1}%
x_{1}^{\prime}$ ($z_{2}x_{2}^{\prime}$ and $x_{2}z_{2}^{\prime}$)
simultaneously, and thus gives also the readout of $z_{1}z_{1}^{\prime}\cdot
x_{1}x_{1}^{\prime}$ ($z_{2}x_{2}^{\prime}\cdot x_{2}z_{2}^{\prime}$). It is
worthwhile to note that apparatus $2$(e) can be replaced with the apparatuses
discriminating the four Bell states in Eq. (\ref{bspp}). For apparatus $2$(f),
the situation is similar.

To summarize, we have demonstrated an all-versus-nothing nonlocality for two
photons, which are maximally entangled both in polarization and in path
degrees of freedom. Since the required measurement of local operators can be
implemented with linear optics, our two-photon proposal is well within the
reach of current quantum optical technology. Note that, untill now, there is
only one experiment \cite{Pan-GHZ} performed to test the GHZ nonlocality, a
kind of all-versus-nothing nonlocality. In this respect, the feasible
experimental scheme, as we suggested in this work, is highly desirable.

Finally, it is interesting to point out that using a single photon as a
two-qubit (polarization and path qubits) system may find important
applications in other contexts, e.g., testing the Kochen-Specker theorem
\cite{KS} with single particles \cite{Simon,Michler}, quantum computing
\cite{QC-pp}, quantum cryptography \cite{cryp-pp}, and entanglement
purification \cite{Simon-Pan}. A recent study shows that the path-polarization
entangled two-photon states (\ref{pp}) can also be used to implement
deterministic and efficient quantum cryptography \cite{Zhao}.

We thank Jiang-Feng Du, Zhi Zhao and Marek Zukowski for useful discussions.
This work was supported by the National Natural Science Foundation of China,
the Chinese Academy of Sciences and the National Fundamental Research Program
(under Grant No. 2001CB309300). This work was also supported by the Austrian
Science Foundation FWF, Project No. F1506.

\end{document}